\documentclass[reprint,notitlepage,showpacs,preprintnumbers,nofootinbib,amsmath,amssymb,aps,prd,floatfix,superscriptaddress]{revtex4-1}
\pdfoutput=1
\usepackage{graphicx}
\usepackage{bm}
\usepackage{subfigure}
\usepackage{url}
\usepackage[hyperindex]{hyperref}
\usepackage[svgnames]{xcolor}
\usepackage[ddmmyy,24hr]{datetime}
\usepackage{bigdelim}
\usepackage{booktabs}
\usepackage{dcolumn}
\usepackage{multirow}
\usepackage{subfigure}
\usepackage{cancel}
\usepackage{stackrel}
\usepackage{paralist}
\usepackage{xspace}
\usepackage{slashed}
\newcommand{\nua}[1]{\ensuremath{\rlap{\kern-2.5pt\ensuremath{\overset{\scriptscriptstyle(-)}{\phantom{\nu}}}}{\ensuremath{{\nu}_{#1}}}}}

\usepackage{enumitem}

\begin{document}

\preprint{CERN-TH-2019-208}

\title{A new analysis of the MiniBooNE low-energy excess}

\author{C. Giunti}
\email{carlo.giunti@to.infn.it}
\affiliation{Istituto Nazionale di Fisica Nucleare (INFN), Sezione di Torino, Via P. Giuria 1, I--10125 Torino, Italy}

\author{A. Ioannisian}
\email{ara.ioannisyan@cern.ch}
\affiliation{CERN, Theory Division, CH-1211 Geneva 23, Switzerland}
\affiliation{Yerevan Physics Institute, Alikhanian Brothers 2, Yerevan-36, Armenia}
\affiliation{Institute for Theoretical Physics and Modeling, Yerevan-36, Armenia}

\author{G. Ranucci}
\email{gioacchino.ranucci@mi.infn.it}
\affiliation{Istituto Nazionale di Fisica Nucleare (INFN), Sezione di Milano, I--20133 Milano, Italy}

\date{19 December 2020}

\begin{abstract}
\hfill\textbf{Corrected version of JHEP 11 (2020) 146}\hfill\null
\\
We present the results of a new analysis of the data of the MiniBooNE experiment
taking into account the additional background of photons from $\Delta^{+/0}$ decay
proposed in Ref.~\cite{Ioannisian:2019kse}
and
additional contributions due to
coherent photon emission,
incoherent production of higher mass resonances,
and incoherent non-resonant nucleon production.
We show that the new background can explain part of the MiniBooNE
low-energy excess and
the statistical significance of the MiniBooNE indication in favor of
short-baseline neutrino oscillation
decreases from
$5.1\sigma$
to
$3.6\sigma$.
We also consider the implications
for short-baseline neutrino oscillations in the 3+1 active-sterile neutrino mixing framework.
We show that the new analysis of the MiniBooNE data
indicates smaller active-sterile neutrino mixing
and may lead us towards a solution of the appearance-disappearance tension
in the global fit of short-baseline neutrino oscillation data.
\end{abstract}


\maketitle

\section{Introduction}
\label{sec:intro}

The MiniBooNE experiment~\cite{Aguilar-Arevalo:2018gpe} found a significant excess of low-energy
$\nu_{e}$-like events that could be due to short-baseline $\nu_{\mu}\to\nu_{e}$ oscillations
generated by active-sterile neutrino mixing~\cite{Giunti:2019aiy,Diaz:2019fwt,Boser:2019rta}
or to other physics beyond the Standard Model~\cite{Bertuzzo:2018itn,Ballett:2018ynz,Arguelles:2018mtc,Coloma:2019qqj}.
However,
the oscillation explanation of the MiniBooNE low-energy excess
is in tension with the data of other short-baseline neutrino oscillation
experiments~\cite{Giunti:2013aea,Ericson:2016yjn,Gariazzo:2017fdh}
and the non-oscillatory explanations are disfavored by other
measurements~\cite{2007.14411}.
A possible solution of this conundrum lies in a reevaluation of the
estimated background that can decrease the low-energy excess.
Among the different sources of background an important one is the
single-$\gamma$ background that cannot be distinguished from
$\nu_{e}$-like events in the MiniBooNE detector.

The MiniBooNE single-$\gamma$ background
was studied theoretically in Ref.~\cite{Hill:2010zy},
where it was found that it is a factor of about 2 larger
than that estimated by the MiniBooNE collaboration.
On the other hand,
the later theoretical studies in Refs.~\cite{Zhang:2012xn,Wang:2014nat}
found an approximate agreement with the MiniBooNE estimate.
However,
doubts on the real value of the MiniBooNE single-$\gamma$ background still remain.
They motivated the ongoing investigation in the
MicroBooNE
experiment at Fermilab
\cite{Gollapinni:2015lca},
which is able to distinguish between photon and $\nu_{e}$
events by using a Liquid Argon Time Projection Chamber (LArTPC).
If the MicroBooNE experiment finds an excess of single-$\gamma$
events that can explain the MiniBooNE low-energy excess,
it will be necessary to understand its origin.

In this paper we propose an increase of the estimated
MiniBooNE single-$\gamma$ background
that can explain part of the MiniBooNE
low-energy excess.
In Section~\ref{sec:bck}
we present the reasons of the increase of the MiniBooNE single-$\gamma$ background
and we calculate the enhancement.
In Section~\ref{sec:sbl} we discuss the implications for short-baseline active-sterile oscillations,
and in Section~\ref{sec:conclusions} we draw our conclusions.

\section{The MiniBooNE single-$\gamma$ background}
\label{sec:bck}

In MiniBooNE the single-$\gamma$ background is
thought to be
due mainly to
$\Delta \to N \gamma$ photons produced by the decay
$\Delta^{+/0} \to p/n + \gamma$
of $\Delta^{+/0}$'s produced in neutral-current $\nu_{\mu}$ interactions with the
mineral oil ($\text{C}\text{H}_{2}$) of the detector.
The MiniBooNE collaboration estimated this background
through the measurement of $\pi^{0}$'s that are produced by the decay
$\Delta^{+/0} \to p/n + \pi^{0}$,
using the branching fractions~\cite{Ignarra:2014yqa,Tanabashi:2018oca}
\begin{align}
\null & \null
\text{Br}( \Delta^{+/0} \to p/n + \gamma )
=
( 6.0 \pm 0.5 ) \times 10^{-3}
,
\label{Brg}
\\
\null & \null
\text{Br}( \Delta^{+/0} \to p/n + \pi^{0} )
\simeq
2/3
.
\label{Brp}
\end{align}
Final state interactions (FSI) cause the absorption of a fraction of the $\pi^{0}$'s
in the carbon nucleus that was estimated to be about 37.5\%
by the MiniBooNE collaboration~\cite{Aguilar-Arevalo:2020nvw}\footnote{
In this paper we do not consider the new MiniBooNE data presented in
Ref.~\cite{Aguilar-Arevalo:2020nvw}
because there is still no available data release.
}.
However,
in Ref.~\cite{Ioannisian:2019kse} one of us noted that
measurements of $\pi^{0}$ photoproduction on nuclei~\cite{Krusche:2004uw,Krusche:2004xz}
indicate that the fraction of $\pi^{0}$'s that emerge from the nucleus
and can be observed is given by
\begin{equation}
\dfrac{ N_{\pi^{0}}^{\text{FSI}} }{ N_{\pi^{0}}^{0} }
=
\dfrac{ \sigma_{\text{FSI}}( \gamma + {}^{A}\mathcal{N} \to \pi^{0} + X ) }{ \sigma_{0}( \gamma + {}^{A}\mathcal{N} \to \pi^{0} + X ) }
\simeq
\dfrac{ A^{2/3} }{ A }
=
A^{-1/3}
,
\label{Rpi0}
\end{equation}
where $A$ is the mass number of the target nucleus ${^A}\mathcal{N}$,
$\sigma_{\text{FSI}}$ denotes the measured cross section which includes final state interactions
and
$\sigma_{0}$ denotes the theoretical cross section without final state interactions.
Since the nuclear radius scales approximately as $A^{1/3}$,
the $A^{2/3}$ dependency of
$\sigma_{\text{FSI}}( \gamma + {}^{A}\mathcal{N} \to \pi^{0} + X )$
indicates that only the nuclear surface contributes,
whereas all the $\pi^{0}$ that are produced in the nuclear interior are absorbed~\cite{Krusche:2004zc}.
However,
the uncertainties of the $A^{2/3}$ scaling in Eq.~(\ref{Rpi0}) are not known.

Our proposal is to consider the extreme total absorption
of the $\pi^{0}$ that are produced in the nuclear interior
as indicated by the $\pi^{0}$ photoproduction data~\cite{Krusche:2004uw,Krusche:2004xz}
as the possible cause of an increase of the estimated MiniBooNE single-$\gamma$ background
that can explain at least part of the MiniBooNE low-energy excess.
In other words, we consider Eq.~(\ref{Rpi0}) as an \emph{ansatz}
of the effects of FSI $\pi^{0}$ absorption in a nucleus
that is motivated by the photoproduction data.
The resulting estimate of
the probability of $\pi^0$ escape from the $^{12}\text{C}$ nucleus is
\begin{equation}
\widetilde{S}_{\text{C}}(\pi^0)
\simeq
12^{-1/3}
=
0.437
,
\label{SGIR}
\end{equation}
that is smaller than that estimated
by the MiniBooNE collaboration~\cite{Aguilar-Arevalo:2020nvw},
\begin{equation}
S_{\text{C}}^{\text{MB}}(\pi^0)
=
0.625
.
\label{SMB}
\end{equation}
According to our estimation,
the number of $\Delta^{+/0}$ produced in neutral-current $\nu_{\mu}$ interactions with $^{12}\text{C}$
and the number of $\gamma$'s generated by their decay is a factor
$[\widetilde{S}_{\text{C}}(\pi^0)]^{-1} \simeq 2.3$
larger than that obtained from the measurement of $\pi^{0}$'s without taking into account FSI.
This enhancement of the $\Delta \to N \gamma$
background due to $\pi^{0}$ FSI in the $^{12}\text{C}$ nucleus
is in approximate agreement with the
theoretical estimation of a factor about 2.4 in Ref.~\cite{Leitner:2008fg}
and it is larger than the factor
$[S_{\text{C}}^{\text{MB}}(\pi^0)]^{-1}=1.6$
considered by the MiniBooNE collaboration~\cite{Aguilar-Arevalo:2020nvw}.

\begin{table*}[t!]
\renewcommand{\arraystretch}{1.45}
\begin{center}
\begin{tabular}{c|ccc|ccc}
\hline\hline
& \multicolumn{3}{c|}{$\nu$ mode} & \multicolumn{3}{c}{$\bar \nu$ mode}\\
$E^\mathrm{QE}_{\nu}$(GeV) & [0.2,0.3] & [0.3,0.475] & [0.475,1.3] & [0.2,0.3] & [0.3,0.475] & [0.475,1.3]\\
\hline
$f^{\text{th}}_{\text{coh}}$ & 0.09 & 0.13 & 0.06 & 0.16 & 0.16 & 0.07 \\
$f^{\text{th}}_{N^{*}}$ & 0.02 & 0.02 & 0.13 & 0.03 & 0.02 & 0.13 \\
$\widetilde{R}/R_{\text{MB}}$ & 1.52 & 1.56 & 1.61 & 1.62 & 1.61 & 1.62 \\
\hline\hline
\end{tabular}
\end{center} 
\caption{ \label{tab:enhancement}
Estimations of
$f^{\text{th}}_{\text{coh}}$ and $f^{\text{th}}_{N^{*}}$
from Table~2 of Ref.~\cite{Wang:2014nat}
in three ranges of reconstructed neutrino energy $E^\mathrm{QE}_\nu$
in the $\nu$ and $\bar\nu$ modes of the MiniBooNE experiment,
and the corresponding
values of the enhancement factor $\widetilde{R}/R_{\text{MB}}$
that we obtained considering $f^{\text{th}}_{N}\simeq0.1$~\cite{Zhang:2012xn}
and our value (\ref{SGIR}) of
the probability of $\pi^0$ escape from the $^{12}\text{C}$ nucleus.
}
\end{table*}

Moreover,
the MiniBooNE collaboration assumed that
``single gamma events are assumed to come entirely from $\Delta$ radiative
decay''~\cite{Aguilar-Arevalo:2020nvw},
neglecting the additional
contributions to $\gamma$ production from
coherent photon emission,
incoherent production of higher mass resonances,
and incoherent non-resonant nucleon production~\cite{Zhang:2012xn,Wang:2014nat}.
Taking into account these contributions and our value (\ref{SGIR}) of
the probability of $\pi^0$ escape from the $^{12}\text{C}$ nucleus,
the ratio of single-$\gamma$ events to NC $\pi^{0}$ events is given by
\begin{widetext}
\begin{equation}
\widetilde{R}
=
\dfrac
{
N_{\text{H}}^{\text{th}}(\Delta \to N \gamma)
+
N_{\text{C}}^{\text{th}}(\Delta \to N \gamma)
+
N_{\text{coh}}^{\text{th}}(\gamma)
+
N^{\text{th}}(N^{*} \to N \gamma)
+
N^{\text{th}}(N \to N \gamma)
}
{\widetilde{N}_{\text{tot}}^{\text{th,obs}}(\pi^0)}
.
\label{Rt1}
\end{equation}
\end{widetext}
The contributions in the numerator are, respectively,
the theoretically predicted numbers of single-$\gamma$ events due to
$\Delta \to N \gamma$ in $\text{H}$,
$\Delta \to N \gamma$ in $\text{C}$,
coherent photon emission,
incoherent production of higher mass resonances
($N^{*} \to N \gamma$)
and incoherent non-resonant nucleon production
($N \to N \gamma$).
The denominator is the theoretically predicted total number of observed $\pi^0$ events.
Note that only the denominator of Eq.~(\ref{Rt1})
depends on the probability of $\pi^0$ escape from the $^{12}\text{C}$ nucleus,
because a larger escape probability implies a larger number of observed $\pi^0$ events.
The tilde notation indicates that $\widetilde{N}_{\text{tot}}^{\text{th,obs}}(\pi^0)$
corresponds to our value $\widetilde{S}_{\text{C}}(\pi^0)$
in Eq.~(\ref{SGIR}) of such probability.

We can write Eq.~(\ref{Rt1}) as
\begin{align}
\widetilde{R}
=
\null & \null
\dfrac
{
N_{\text{H}}^{\text{th}}(\Delta \to N \gamma)
+
N_{\text{C}}^{\text{th}}(\Delta \to N \gamma)
}
{N_{\text{tot}}^{\text{th,obs}}(\pi^0)}
\left(
\dfrac
{N_{\text{tot}}^{\text{th,obs}}(\pi^0)}
{\widetilde{N}_{\text{tot}}^{\text{th,obs}}(\pi^0)}
\right)
\nonumber
\\
\null & \null
\times
\left( 1 + f^{\text{th}}_{\text{coh}} + f^{\text{th}}_{N^{*}} + f^{\text{th}}_{N} \right)
,
\label{Rt2}
\end{align}
where
$N_{\text{tot}}^{\text{th,obs}}(\pi^0)$
is the total number of observed $\pi^0$ events estimated by the MiniBooNE collaboration
using the probability of $\pi^0$ escape from the $^{12}\text{C}$ nucleus
$S_{\text{C}}^{\text{MB}}(\pi^0)$ in Eq.~(\ref{SMB}).
In Eq.~(\ref{Rt2})
$f^{\text{th}}_{\text{coh}}$,
$f^{\text{th}}_{N^{*}}$, and
$f^{\text{th}}_{N}$
are, respectively,
the theoretically predicted ratios of $\gamma$'s generated
coherently, by higher mass resonances, and non-resonant nucleon production
with respect to those generated by $\Delta$ decay.
In this way, we separated these contributions from the
those generated by $\Delta$ decay that were considered by the MiniBooNE collaboration.

The first fraction on the right-hand side of Eq.~(\ref{Rt2})
is the ratio of single-$\gamma$ events to NC $\pi^{0}$ events
estimated by the MiniBooNE collaboration:
\begin{equation}
R_{\text{MB}}
=
\dfrac
{
N_{\text{H}}^{\text{th}}(\Delta \to N \gamma)
+
N_{\text{C}}^{\text{th}}(\Delta \to N \gamma)
}
{N_{\text{tot}}^{\text{th,obs}}(\pi^0)}
=
0.0091
.
\label{RMB}
\end{equation}

The second fraction on the right-hand side of Eq.~(\ref{Rt2})
can be calculated by writing it as
\begin{equation}
\dfrac
{N_{\text{tot}}^{\text{th,obs}}(\pi^0)}
{\widetilde{N}_{\text{tot}}^{\text{th,obs}}(\pi^0)}
=
\dfrac
{N_{\text{abs}}^{\text{th,obs}}(\pi^0)+N_{\text{noabs}}^{\text{th,obs}}(\pi^0)}
{\widetilde{N}_{\text{abs}}^{\text{th,obs}}(\pi^0)+N_{\text{noabs}}^{\text{th,obs}}(\pi^0)}
,
\label{Nr1}
\end{equation}
where
$N_{\text{abs}}^{\text{th,obs}}(\pi^0)$
and
$\widetilde{N}_{\text{abs}}^{\text{th,obs}}(\pi^0)$
are the theoretically predicted numbers of observed $\pi^0$ produced in processes
with absorption of $\pi^0$ in the C nucleus,
whereas
$N_{\text{noabs}}^{\text{th,obs}}(\pi^0)$
is the theoretically predicted numbers of observed $\pi^0$ produced in processes
without absorption of $\pi^0$ in the C nucleus.
Note that only
$N_{\text{abs}}^{\text{th,obs}}(\pi^0)$
and
$\widetilde{N}_{\text{abs}}^{\text{th,obs}}(\pi^0)$
depend on the probability of $\pi^0$ escape from the $^{12}\text{C}$ nucleus
and are given by
$
N_{\text{abs}}^{\text{th,obs}}(\pi^0)
=
N_{\text{abs}}^{\text{th,prod}}(\pi^0)
S_{\text{C}}^{\text{th}}(\pi^0)
$
and
$
\widetilde{N}_{\text{abs}}^{\text{th,obs}}(\pi^0)
=
N_{\text{abs}}^{\text{th,prod}}(\pi^0)
\widetilde{S}_{\text{C}}^{\text{th}}(\pi^0)
$.
Therefore,
we can write Eq.~(\ref{Nr1})
as
\begin{equation}
\dfrac
{N_{\text{tot}}^{\text{th,obs}}(\pi^0)}
{\widetilde{N}_{\text{tot}}^{\text{th,obs}}(\pi^0)}
=
\dfrac
{
1
+
\dfrac{N_{\text{noabs}}^{\text{th,obs}}(\pi^0)}{N_{\text{abs}}^{\text{th,obs}}(\pi^0)}
}
{
\dfrac{\widetilde{S}_{\text{C}}^{\text{th}}(\pi^0)}{S_{\text{C}}^{\text{th}}(\pi^0)}
+
\dfrac{N_{\text{noabs}}^{\text{th,obs}}(\pi^0)}{N_{\text{abs}}^{\text{th,obs}}(\pi^0)}
}
.
\label{Nr2}
\end{equation}
We obtained the value of
$
N_{\text{noabs}}^{\text{th,obs}}(\pi^0)
/
N_{\text{abs}}^{\text{th,obs}}(\pi^0)
$
from the contributions to the MiniBooNE $\pi^{0}$
event sample given in Ref.~\cite{Aguilar-Arevalo:2020nvw}:
\begin{equation}
\dfrac
{N_{\text{noabs}}^{\text{th,obs}}(\pi^0)}
{N_{\text{abs}}^{\text{th,obs}}(\pi^0)}
\simeq
0.496
.
\label{Nr3}
\end{equation}
The resulting enhancement factor of the MiniBooNE single-$\gamma$ background is
\begin{equation}
\dfrac{\widetilde{R}}{R_{\text{MB}}}
\simeq
1.25
\left( 1 + f^{\text{th}}_{\text{coh}} + f^{\text{th}}_{N^{*}} + f^{\text{th}}_{N} \right)
.
\label{Rt3}
\end{equation}

The authors of Ref.~\cite{Wang:2014nat} calculated the
number of single photon events from neutral current interactions at MiniBooNE.
From their Table~2 we obtained the estimates of
$f^{\text{th}}_{\text{coh}}$ and $f^{\text{th}}_{N^{*}}$
in Table~\ref{tab:enhancement},
considering three ranges of $E^\mathrm{QE}_\nu$
in the $\nu$ and $\bar\nu$ modes of the MiniBooNE experiment.
For $f^{\text{th}}_{N}$ we considered the 10\% value estimated in Ref.~\cite{Zhang:2012xn}.
Note that we did not use the absolute values of the events calculated in
Refs.~\cite{Zhang:2012xn,Wang:2014nat},
that are in agreement with the MiniBooNE estimates,
and hence in disagreement with our estimations.
We used only the relative values of the $\gamma$'s generated
coherently, by higher mass resonances, and non-resonant nucleon production,
whose estimation can be considered more accurate.

\begin{figure*}[!t]
\centering
\setlength{\tabcolsep}{0pt}
\begin{tabular}{cc}
\subfigure[]{\label{fig:hst-mbn-bck}
\includegraphics*[width=0.45\linewidth]{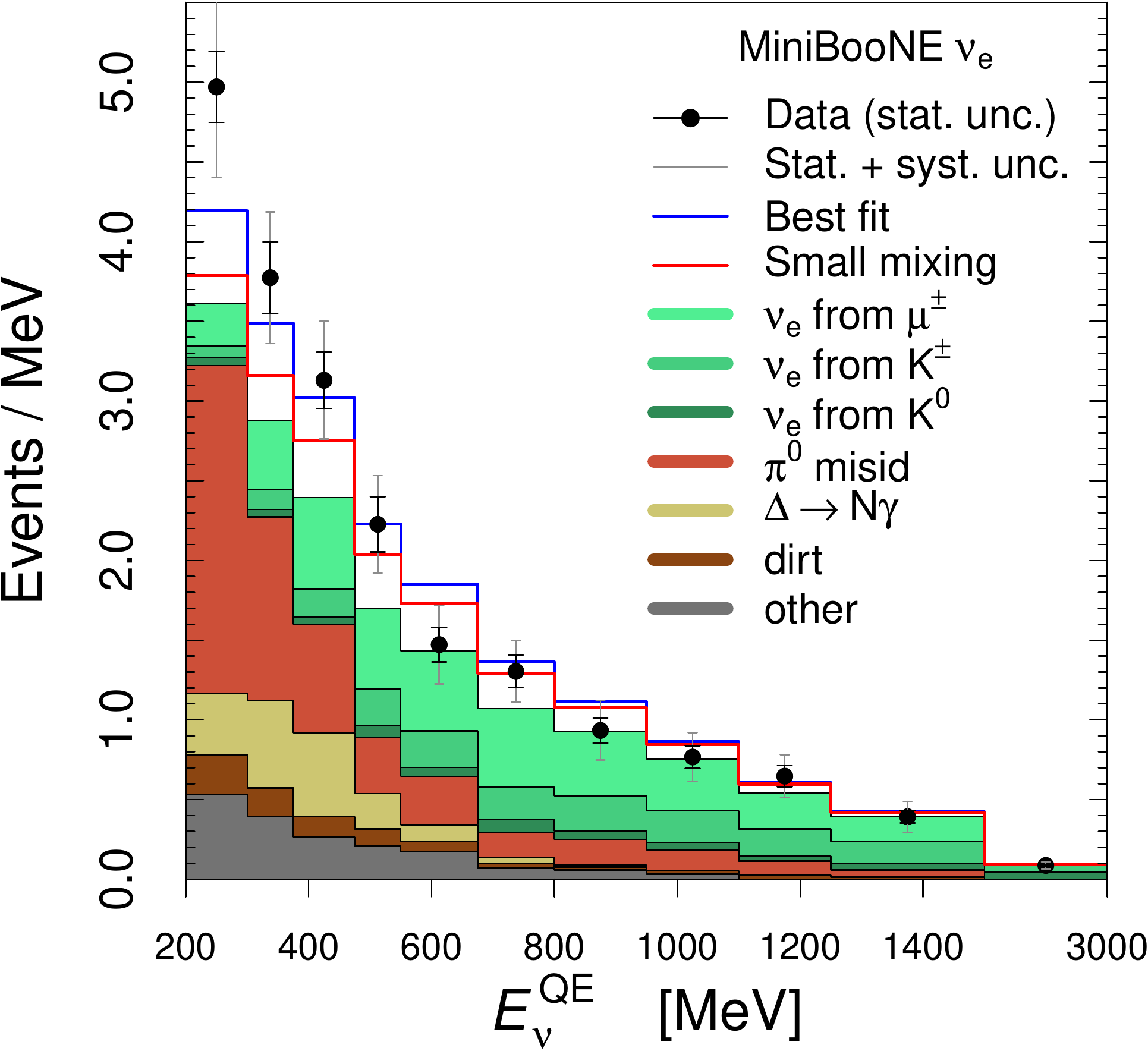}
}
&
\subfigure[]{\label{fig:hst-mba-bck}
\includegraphics*[width=0.45\linewidth]{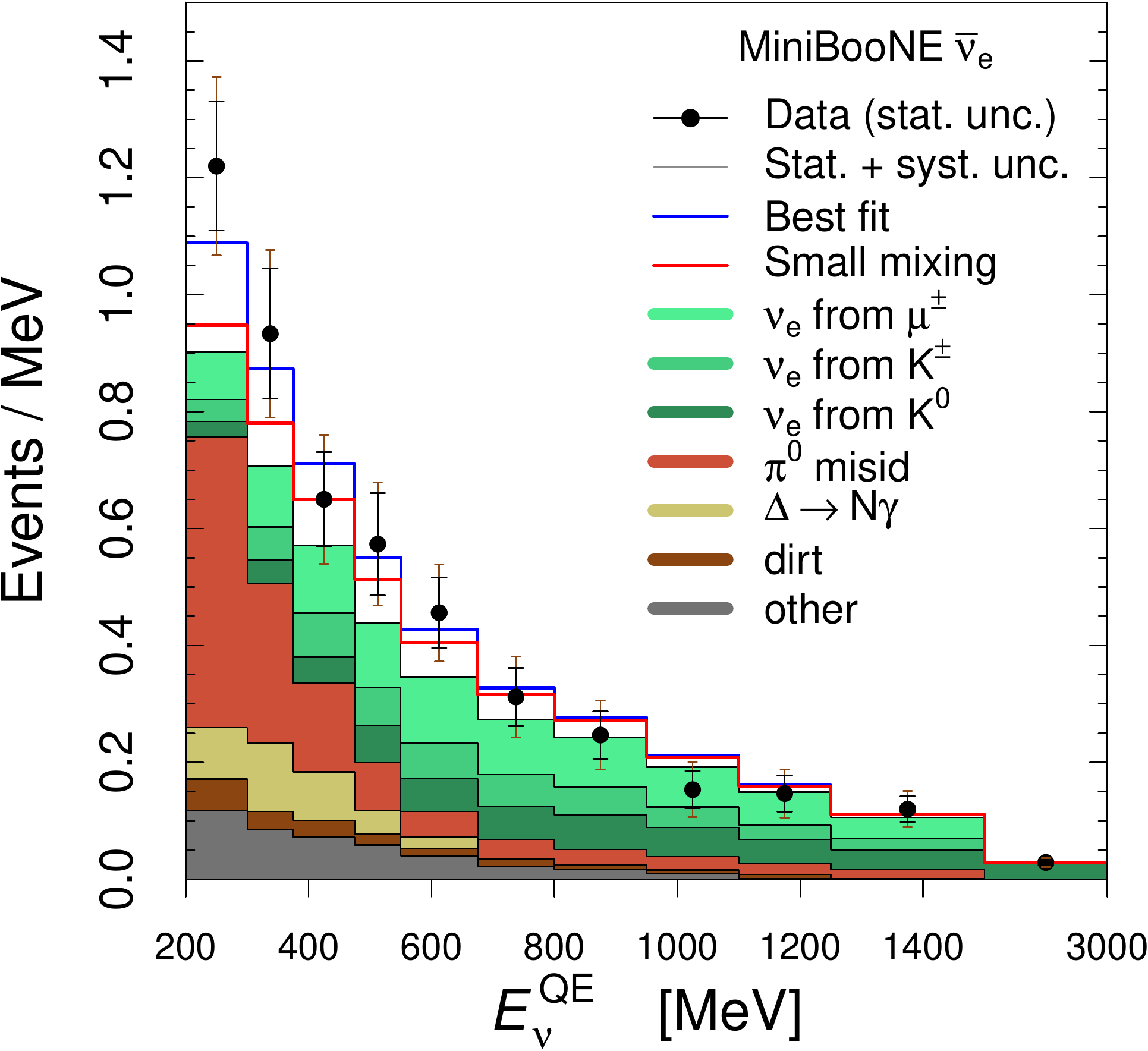}
}
\\
\subfigure[]{\label{fig:hst-mbn-bck+fsi}
\includegraphics*[width=0.45\linewidth]{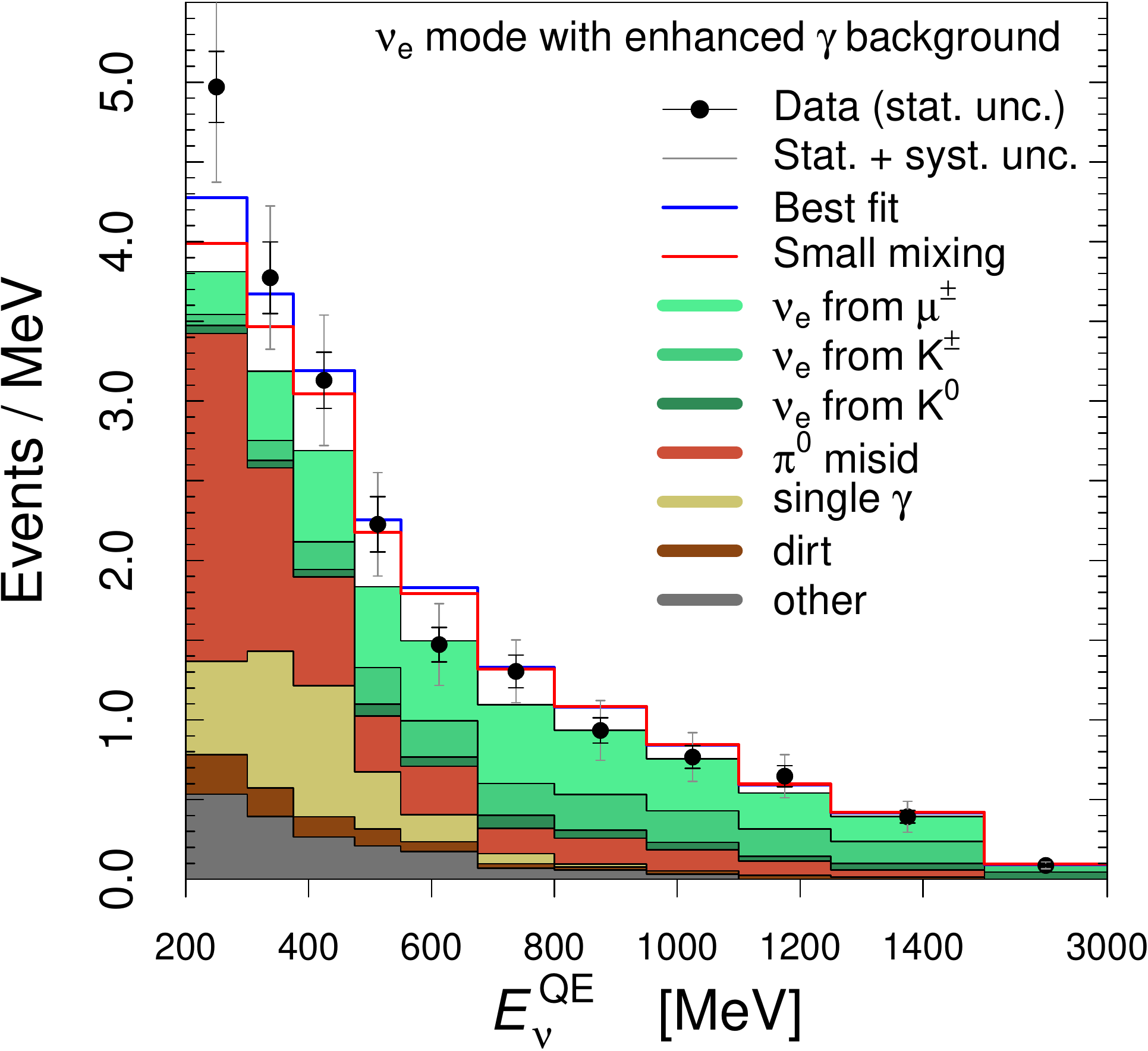}
}
&
\subfigure[]{\label{fig:hst-mba-bck+fsi}
\includegraphics*[width=0.45\linewidth]{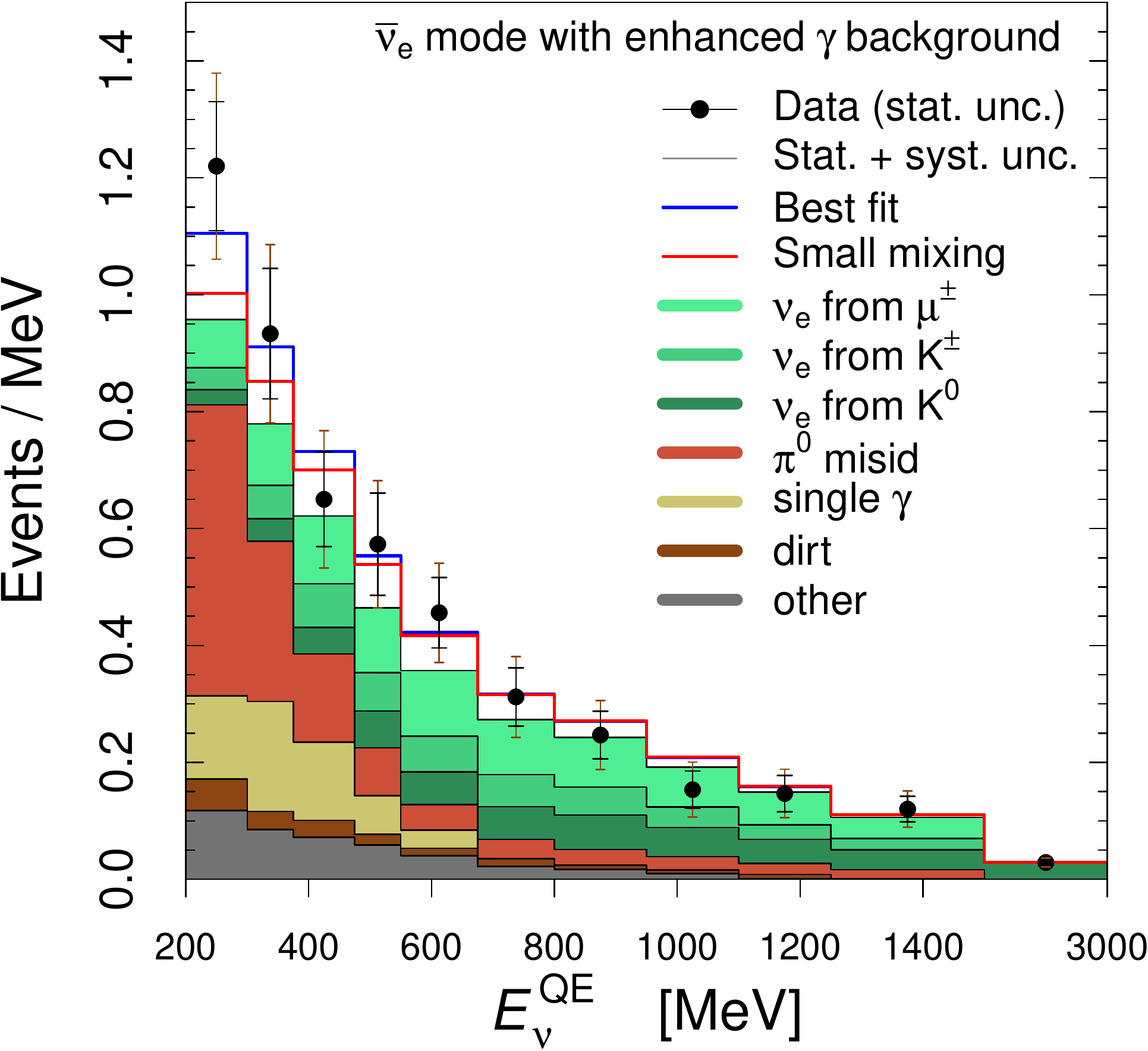}
}
\end{tabular}
\caption{ \label{fig:hst}
Comparison of a reproduction of the MiniBooNE event histograms in
\subref{fig:hst-mbn-bck} neutrino and
\subref{fig:hst-mba-bck} antineutrino mode
from Refs.~\cite{Aguilar-Arevalo:2018gpe,Aguilar-Arevalo:2013pmq}
with our versions \subref{fig:hst-mbn-bck+fsi} and \subref{fig:hst-mba-bck+fsi}
obtained with the enhanced single-$\gamma$ background
due to $A^{1/3}$ $\pi^{0}$ FSI in $^{12}\text{C}$,
coherent photon emission,
incoherent production of higher mass resonances,
and incoherent non-resonant nucleon production.
The blue and red lines show, respectively, the expectations for neutrino oscillations corresponding
to the best fit in Table~\ref{tab:fit} (almost maximal mixing) and
the case of small mixing with
$\sin^2\!2\vartheta_{e\mu} = 2.5 \times 10^{-3}$
and
$\Delta{m}^2_{41} = 0.8 \, \text{eV}^2$.
}
\end{figure*}

As shown in Table~\ref{tab:enhancement},
we find an enhancement
$\widetilde{R}/R_{\text{MB}}$
of the single-$\gamma$ background in MiniBooNE
by a factor between 1.52 and 1.62 depending on the
energy range and neutrino or antineutrino mode of the MiniBooNE experiment.
This increase of the single-$\gamma$ background
can explain in part the low-energy MiniBooNE excess,
because its largest contribution occur in the lowest energy bins,
as one can see from Figures~\ref{fig:hst-mbn-bck} and~\ref{fig:hst-mba-bck}
that reproduce the MiniBooNE event histograms in
neutrino and antineutrino mode
in Refs.~\cite{Aguilar-Arevalo:2018gpe,Aguilar-Arevalo:2013pmq}.

Figure~\ref{fig:hst} shows a comparison of the standard
MiniBooNE event histograms
(Figures~\ref{fig:hst-mbn-bck} and~\ref{fig:hst-mba-bck})
with those obtained with our reevaluation of the single-$\gamma$ background
(Figures~\ref{fig:hst-mbn-bck+fsi} and~\ref{fig:hst-mba-bck+fsi}).
One can see that in the reproductions
\ref{fig:hst-mbn-bck} and \ref{fig:hst-mba-bck}
of the original MiniBooNE histograms
the low-energy bins show a large excess with respect to the background prediction.
The excess is significantly reduced
with our enhanced single-$\gamma$ background.
Only the first energy bin remains with a large visible excess.

The improvement of the fit of the MiniBooNE data is quantified by
$\chi^2/\text{NDF} = 36.9 / 22$,
corresponding to a goodness-of-fit of $ 2\%$
obtained with the enhanced single-$\gamma$ background,
compared to
$\chi^2/\text{NDF} = 53.0 / 22$,
corresponding to a goodness-of-fit of $0.02\%$,
obtained in the standard analysis of MiniBooNE data.

\section{Short-baseline neutrino oscillations}
\label{sec:sbl}

The reevaluation of the low-energy MiniBooNE excess has important implications
for the interpretation of the MiniBooNE data
in terms of short-baseline neutrino oscillations due to active-sterile neutrino mixing.
In the following we consider the 3+1 scenario in which in addition to the three standard light massive neutrinos
$\nu_{1}$, $\nu_{2}$, $\nu_{3}$,
with respective masses
$m_{1}$, $m_{3}$, $m_{3}$,
there is a heavier neutrino $\nu_{4}$ with mass $m_{4}$.
The masses of the three standard light massive neutrinos
have small separations, determined by the measurements of solar, atmospheric and long-baseline oscillations:
$ \Delta{m}^2_{21} \simeq 7.4 \times 10^{-5} \, \text{eV}^2 $
and
$ |\Delta{m}^2_{31}| \simeq 2.5 \times 10^{-3} \, \text{eV}^2 $,
with $ \Delta{m}^2_{ij} \equiv m_{i}^2 - m_{j}^2 $.
A much larger squared mass difference
$ \Delta{m}^2_{41} \simeq \Delta{m}^2_{42} \simeq \Delta{m}^2_{43} \gtrsim 0.1 \, \text{eV}^2 $
can generate short-baseline
$\nua{\mu}\to\nua{e}$
oscillations that may explain the LSND and MiniBooNE anomalies,
as well as other indications of short-baseline neutrino oscillations~\cite{Giunti:2019aiy,Diaz:2019fwt,Boser:2019rta}.
The probability of short-baseline
$\nua{\mu}\to\nua{e}$
oscillations is given by
\begin{equation}
P^{\text{SBL}}_{\nua{\mu}\to\nua{e}}
=
\sin^2\!2\vartheta_{e\mu}
\sin^2\left( \dfrac{ \Delta{m}^2_{41} L }{ 4 E } \right)
,
\label{Pme}
\end{equation}
where $E$ is the neutrino energy,
$L$ is the source-detector distance,
and
$ \sin^2\!2\vartheta_{e\mu} = 4 |U_{e4}|^2 |U_{\mu4}|^2 $,
where $U$ is the $4\times4$ unitary mixing matrix.

\begin{figure}[!t]
\centering
\includegraphics*[width=\linewidth]{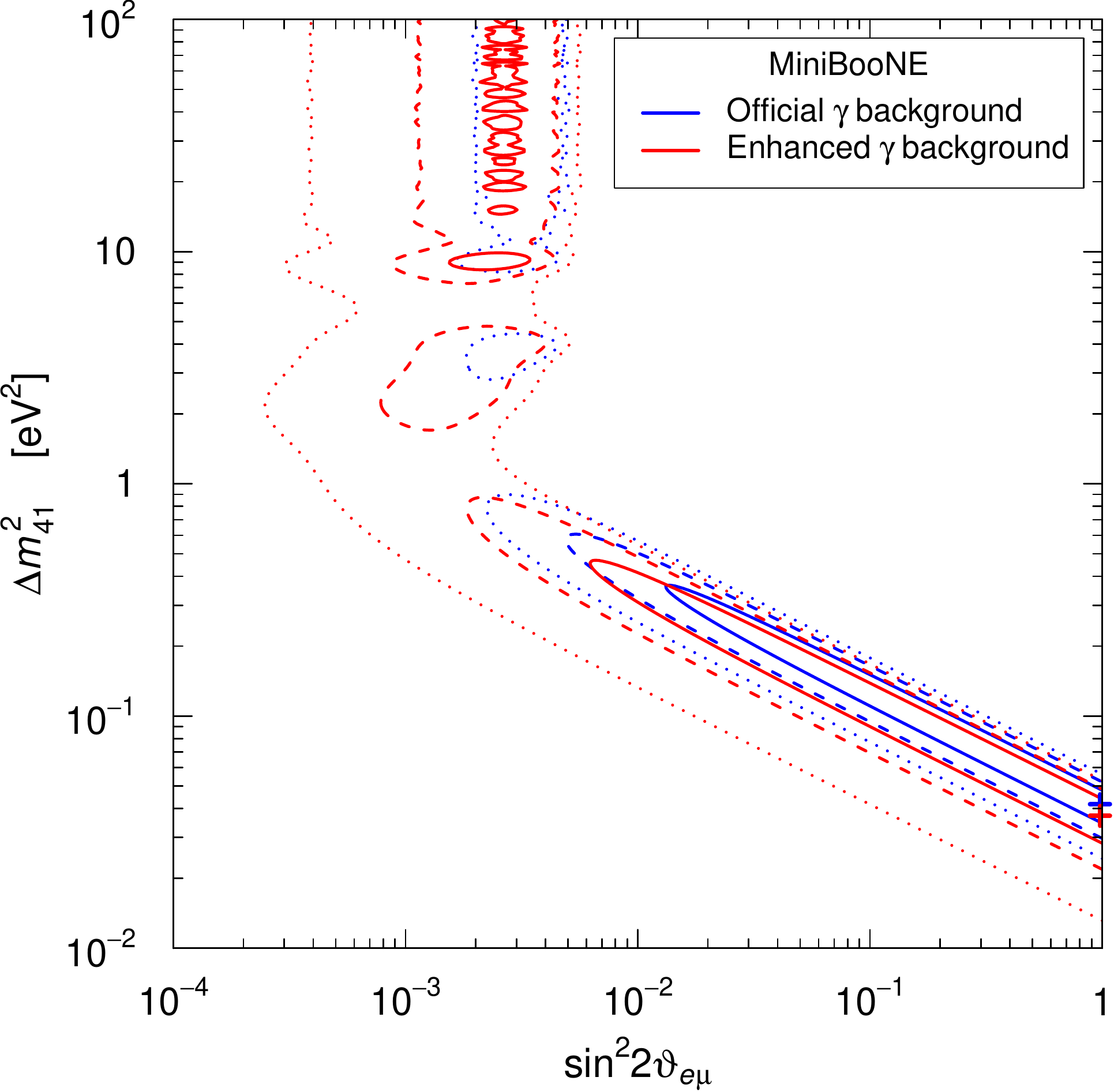}
\caption{ \label{fig:MB+FSI}
Contours enclosing
$1\sigma$ (solid),
$2\sigma$ (dashed), and
$3\sigma$ (dotted)
allowed regions in the
($\sin^2\!2\vartheta_{e\mu},\Delta{m}^2_{41}$)
plane obtained from the analysis of
MiniBooNE data
without (blue) and with (red)
our single-$\gamma$ background enhancement.
}
\end{figure}

Figure~\ref{fig:MB+FSI}
shows a comparison of the standard allowed regions in the
($\sin^2\!2\vartheta_{e\mu},\Delta{m}^2_{41}$)
plane obtained from the analysis of the MiniBooNE data
and those obtained with our enhanced single-$\gamma$ background.
The goodness-of-fit and the best-fit values of the oscillation parameters
are listed in Table~\ref{tab:fit},
and the best fit event histograms are shown in Figure~\ref{fig:hst}.
The difference between the $\chi^2_{\text{min}}$ with oscillations and the $\chi^2$
without oscillations is
$30.2$
and
$16.0$,
without and with our enhanced single-$\gamma$ background,
respectively.
Taking into account that there is a difference of two degrees of freedom corresponding to the two fitted oscillation parameters
$\sin^2\!2\vartheta_{e\mu}$ and $\Delta{m}^2_{41}$,
the statistical significance of the MiniBooNE indication in favor of oscillation
decreases from
$5.1\sigma$
to
$3.6\sigma$
with the introduction of our enhanced single-$\gamma$ background
(the corresponding $\chi^2$ probabilities of the background-only fit
relative to the best oscillation fit are
$2.8 \times 10^{-7}$
and
$3.4 \times 10^{-4}$)\footnote{
In Ref.~\cite{Aguilar-Arevalo:2018gpe}
the MiniBooNE collaboration obtained a probability of the background-only fit
relative to the best oscillation fit of $6 \times 10^{-7}$,
which corresponds to $5.0\sigma$.
The small difference with our result is due to a different analysis
of the data performed by the MiniBooNE collaboration with respect to that recommended in their
data release~\cite{MiniBooNE-DR-1805.12028}.
In particular, they considered only the data below 1250 MeV because that upper limit
``was chosen by the collaboration before unblinding the data in 2007''~\cite{Aguilar-Arevalo:2018gpe}.
We have no reason to implement this restriction.
}.

From Figure~\ref{fig:MB+FSI}
one can see that the allowed regions in the ($\sin^2\!2\vartheta_{e\mu},\Delta{m}^2_{41}$)
plane change significantly by taking into account our enhanced single-$\gamma$ background.
Although the best fit remains at quasi-maximal mixing,
there is an extension of the allowed regions towards small values of the mixing parameter
$\sin^2\!2\vartheta_{e\mu}$.
In particular, the $3\sigma$ allowed region becomes
a band that allows small values of $\sin^2\!2\vartheta_{e\mu}$
of the order of $10^{-3}$
for
$\Delta{m}^2_{41} \gtrsim 0.4 \, \text{eV}^2$.
This is beneficial, because large active-sterile mixing is disfavored by
solar, atmospheric and long-baseline neutrino oscillation data~\cite{Giunti:2019aiy,Diaz:2019fwt,Boser:2019rta}.

The values of the goodness-of-fit in Table~\ref{tab:fit} show that
the fit of MiniBooNE data with the enhanced single-$\gamma$ background
is better than the one without,
although the difference is small.
However, it is more important that
the ``Small mixing'' event histograms in Figure~\ref{fig:hst}
show that the small number of signal events resulting from
the small mixing $\sin^2\!2\vartheta_{e\mu} = 2.5 \times 10^{-3}$
can fit better the low-energy data with our enhanced single-$\gamma$ background.
Only the excess in the first bin is not well fitted.

\begin{table*}[!t]
\centering
\renewcommand{\arraystretch}{1.3}
\begin{tabular}[t]{c|}
\\
\hline
$\chi^{2}_{\text{min}}$\\
NDF\\
GoF\\
$\Delta{m}^{2\text{(bf)}}_{41}$\\
$\sin^22\vartheta_{e\mu}^{\text{(bf)}}$\\
\end{tabular}%
\begin{tabular}[t]{c}
MB\\
\hline
$22.8$\\
$20$\\
$30\%$\\
$0.042$\\
$0.98$\\
\end{tabular}%
\begin{tabular}[t]{c|}
$\widetilde{\text{MB}}$\\
\hline
$20.9$\\
$20$\\
$40\%$\\
$0.037$\\
$0.98$\\
\end{tabular}%
\begin{tabular}[t]{c}
LSND+MB\\
\hline
$23.0$\\
$31$\\
$85\%$\\
$0.042$\\
$1.0$\\
\end{tabular}%
\begin{tabular}[t]{c|}
LSND+$\widetilde{\text{MB}}$\\
\hline
$21.8$\\
$31$\\
$89\%$\\
$0.040$\\
$1.0$\\
\end{tabular}%
\begin{tabular}[t]{c}
App+MB\\
\hline
$76.9$\\
$75$\\
$42\%$\\
$0.58$\\
$0.0062$\\
\end{tabular}%
\begin{tabular}[t]{c}
App+$\widetilde{\text{MB}}$\\
\hline
$73.4$\\
$75$\\
$53\%$\\
$0.69$\\
$0.0040$\\
\end{tabular}%
\caption{ \label{tab:fit}
Minimum $\chi^2$,
number of degrees of freedom (NDF)
and Goodness of Fit (GoF)
of the analyses of the data of short-baseline $\protect\nua{\mu}\to\protect\nua{e}$ experiments
discussed in the text
without (MB) and with ($\widetilde{\text{MB}}$)
our enhanced single-$\gamma$ background in MiniBooNE.
$\Delta{m}^{2\text{(bf)}}_{41}$ and $\sin^22\vartheta_{e\mu}^{\text{(bf)}}$
are the best-fit values of the corresponding oscillation parameters.
}
\end{table*}

\begin{figure}[!t]
\centering
\includegraphics*[width=\linewidth]{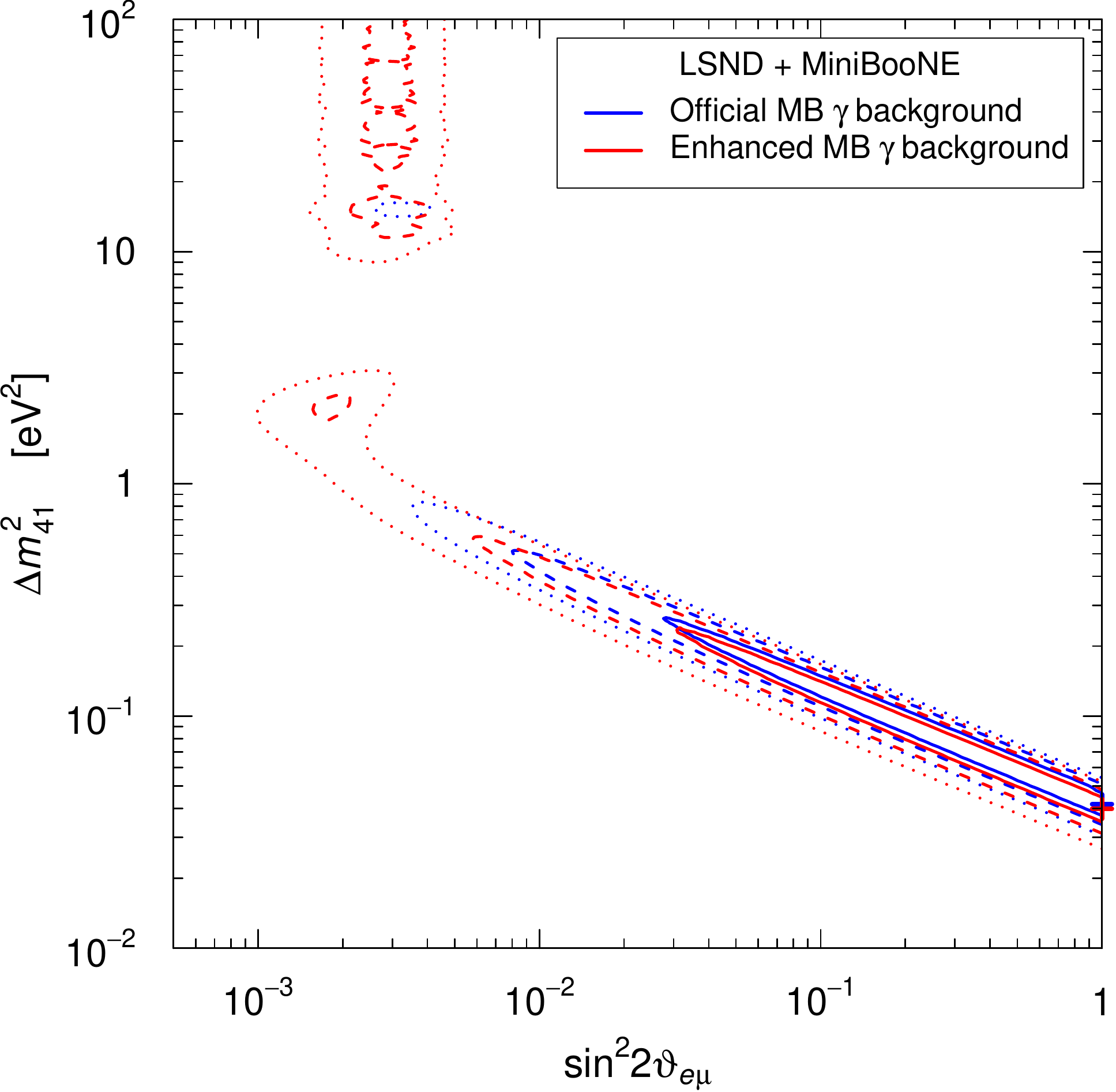}
\caption{ \label{fig:LSND+MB+FSI}
Contours enclosing
$1\sigma$ (solid),
$2\sigma$ (dashed), and
$3\sigma$ (dotted)
allowed regions in the
($\sin^2\!2\vartheta_{e\mu},\Delta{m}^2_{41}$)
plane obtained from the analysis of
LSND and MiniBooNE data
without (blue) and with (red)
the enhanced single-$\gamma$ background.
}
\end{figure}

It is interesting to compare the results of our new fit of the MiniBooNE data
with the indication of the LSND experiment~\cite{Aguilar:2001ty}
in favor of short-baseline $\bar\nu_{\mu}\to\bar\nu_{e}$ oscillations.
Figure~\ref{fig:LSND+MB+FSI}
shows a comparison of the combined LSND and MiniBooNE allowed regions
in the
($\sin^2\!2\vartheta_{e\mu},\Delta{m}^2_{41}$)
plane obtained without and with
the enhanced single-$\gamma$ background.
One can see that the changes are similar to those for MiniBooNE alone
(shown in Figure~\ref{fig:MB+FSI}):
there is a clear shift of the allowed regions towards small mixing.
In particular,
the allowed region with small mixing
around
$\Delta{m}^2_{41} \approx 2 \, \text{eV}^2$
may be compatible with indications of short-baseline $\bar\nu_{e}$
disappearance due to active-sterile neutrino mixing found in
reactor experiments~\cite{Gariazzo:2018mwd,Dentler:2018sju,Diaz:2019fwt,Berryman:2019hme,Giunti:2019fcj,Giunti:2020uhv,Berryman:2020agd}.

\begin{figure}[!t]
\centering
\includegraphics*[width=\linewidth]{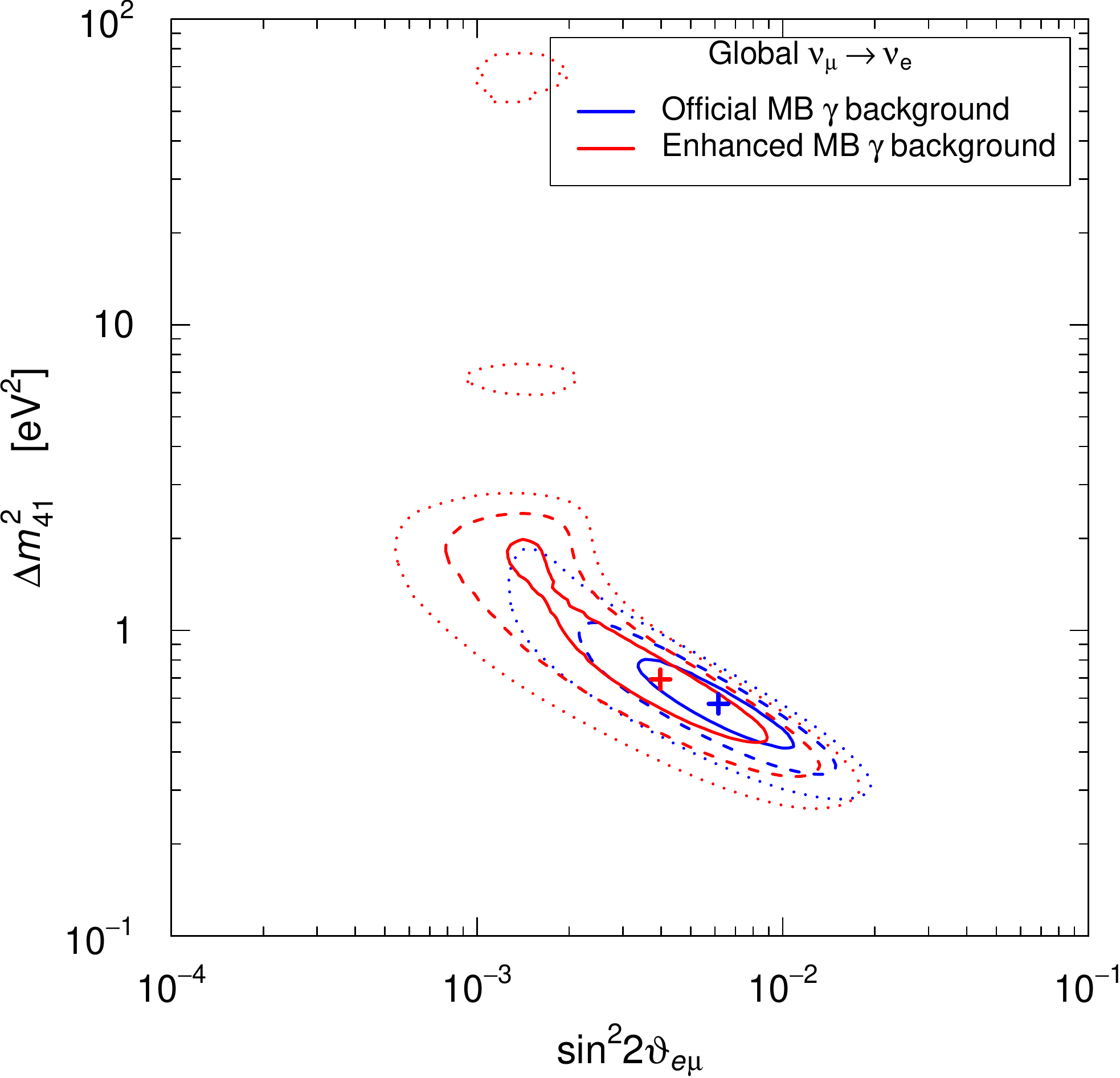}
\caption{ \label{fig:App-MB+FSI}
Contours enclosing
$1\sigma$ (solid),
$2\sigma$ (dashed), and
$3\sigma$ (dotted)
allowed regions in the
($\sin^2\!2\vartheta_{e\mu},\Delta{m}^2_{41}$)
plane obtained from the global analysis of the data of
$\protect\nua{\mu}\to\protect\nua{e}$ oscillation experiments
without (blue) and with (red)
the enhanced single-$\gamma$ background in the analysis of MiniBooNE data.
}
\end{figure}

Finally, we investigated the effects of the analysis of MiniBooNE data
with the enhanced single-$\gamma$ background
on the global fit of the data of short-baseline
$\nua{\mu}\to\nua{e}$
oscillation experiments.
Besides MiniBooNE and LSND,
we considered the data of the
BNL-E776 \cite{Borodovsky:1992pn},
KARMEN \cite{Armbruster:2002mp},
NOMAD \cite{Astier:2003gs},
ICARUS \cite{Antonello:2013gut}
and
OPERA \cite{Agafonova:2013xsk}
experiments,
as done in Ref.~\cite{Gariazzo:2017fdh}.
Figure~\ref{fig:App-MB+FSI}
shows a comparison of the global allowed regions in the
($\sin^2\!2\vartheta_{e\mu},\Delta{m}^2_{41}$)
plane obtained without and with
the enhanced single-$\gamma$ background in the analysis of MiniBooNE data.
One can see that there is again a shift of the allowed regions
towards small mixing.
Although the shift of the best-fit point towards small mixings is not large,
there is a much larger shift of the allowed regions,
that reach $\sin^2\!2\vartheta_{e\mu} \approx 6 \times 10^{-4}$ at $3\sigma$
for
$\Delta{m}^2_{41} \approx 2 \, \text{eV}^2$.
This shift is beneficial for a decrease of the appearance-disappearance tension
in the global fit of short-baseline neutrino oscillation data
in terms of 3+1 active-sterile neutrino mixing~\cite{Okada:1996kw,Bilenky:1996rw,Kopp:2011qd,Giunti:2011gz,Giunti:2011hn,Giunti:2011cp,Conrad:2012qt,Archidiacono:2012ri,Archidiacono:2013xxa,Kopp:2013vaa,Giunti:2013aea,Gariazzo:2015rra,Gariazzo:2017fdh,Dentler:2017tkw,Gariazzo:2018mwd,Dentler:2018sju}.

For example,
the $3\sigma$ upper limit from disappearance data in Figure~7 of Ref.~\cite{Dentler:2018sju}
is about $\sin^2\!2\vartheta_{e\mu} \lesssim 6 \times 10^{-4}$ at $3\sigma$
for
$\Delta{m}^2_{41} \approx 1.3 \, \text{eV}^2$,
that is about the same as the $3\sigma$ lower limit
$\sin^2\!2\vartheta_{e\mu} \gtrsim 6 \times 10^{-4}$
in Figure~\ref{fig:App-MB+FSI}
for the same value of $\Delta{m}^2_{41}$.
It is clear that there is still a considerable appearance-disappearance tension,
but it is significantly smaller than that obtained in Ref.~\cite{Dentler:2018sju}.

The disappearance bound in Figure~5b of Ref.~\cite{Giunti:2019aiy}
is weaker than that obtained in Ref.~\cite{Dentler:2018sju},
with an upper limit
$\sin^2\!2\vartheta_{e\mu} \lesssim 10^{-3}$ at $3\sigma$
for
$\Delta{m}^2_{41} \approx 1.3 \, \text{eV}^2$,
that is compatible with the $3\sigma$ allowed region
in Figure~\ref{fig:App-MB+FSI}.

\section{Conclusions}
\label{sec:conclusions}

In conclusion,
we have shown that a reassessment of the
single-$\gamma$ background from $\Delta^{+/0}$ decay in the MiniBooNE experiment
taking into account the effect of $A^{1/3}$ $\pi^{0}$ FSI proposed in Ref.~\cite{Ioannisian:2019kse}
and additional contributions to the single-$\gamma$ background
can explain in part the low-energy MiniBooNE excess.
In absence of physics beyond the standard three-neutrino mixing,
our enhanced single-$\gamma$ background
leads to a better fit of the data,
with a goodness-of-fit of $ 2\%$,
with respect to the standard analysis,
that has a goodness-of-fit of $0.02\%$.
However,
the MiniBooNE data are still fitted in a better way
considering short-baseline $\nua{\mu}\to\nua{e}$ oscillations
due to active-sterile neutrino mixing,
albeit the statistical significance of the indication in favor of oscillation
decreases from
$5.1\sigma$
to
$3.6\sigma$.
We have shown that in the 3+1 framework
the new analysis of the MiniBooNE data
allows smaller values of active-sterile neutrino mixing
with respect to the standard analysis.
This shift towards small active-sterile neutrino mixing
is beneficial towards a possible solution of the appearance-disappearance tension
in the global fit of short-baseline neutrino oscillation data.

\begin{acknowledgments}
We would like to thank
W.C. Louis and C.A. Ternes
for useful discussions.
The work of C. Giunti was partially supported by the research grant "The Dark Universe: A Synergic Multimessenger Approach" number 2017X7X85K under the program PRIN 2017 funded by the Ministero dell'Istruzione, Universit\`a e della Ricerca (MIUR).
\end{acknowledgments}

%

\end{document}